\documentclass[12pt]{article}
\usepackage{graphicx}
\usepackage{lineno}

\newcommand\pubnumber{SNSN-323-63}
\newcommand\pubdate{\today}

\def\institute{University of Arizona\\
  Tucson, AZ 85719, USA}
\def\support{\footnote{Copyright 2020 CERN for the benefit of the ATLAS and CMS Collaborations. Reproduction of this article or parts of it is allowed as specified in the CC-BY-4.0 license.}}

\def\Title#1{\begin{center} {\Large #1 } \end{center}}
\def\Author#1{\begin{center}{ \sc #1} \end{center}}
\def\Address#1{\begin{center}{ \it #1} \end{center}}

\newcommand\pubblock{\rightline{\begin{tabular}{l} \pubnumber\\
      \pubdate  \end{tabular}}}
\newenvironment{Abstract}{\begin{quotation}  }{\end{quotation}}
\newenvironment{Presented}{\begin{quotation} \begin{center} 
      PRESENTED AT\end{center}\bigskip 
      \begin{center}\begin{large}}{\end{large}\end{center} \end{quotation}}

\def\beq{\begin{equation}}
\def\eeq#1{\label{#1}\end{equation}}
\def\eeqn{\end{equation}}

\def\beqa{\begin{eqnarray}}
\def\eeqa#1{\label{#1}\end{eqnarray}}
\def\eeqan{\end{eqnarray}}

\let\bar=\overbar

\def\Dslash{\not{\hbox{\kern-4pt $D$}}}
\def\dslash{\not{\hbox{\kern-2pt $\del$}}}

\def\msb{{\bar{\ssstyle M \kern -1pt S}}}

\def\TeV{Te\kern -0.1em V}
\def\GeV{Ge\kern -0.1em V}
\newcommand*{\POWHEG}{\textsc{Powheg}}
\newcommand*{\PYTHIA}{\textsc{Pythia}}
\newcommand*{\HERWIG}{\textsc{Herwig}}
\newcommand*{\MadGraph}{\textsc{MadGraph5\_aMC@NLO}}
\newcommand*{\SHERPA}{\textsc{SHERPA}}

\newcommand*{\PW}{\textsc{PW}}
\newcommand*{\HE}{\textsc{H}}
\newcommand*{\Hpp}{\textsc{Hpp}}
\newcommand*{\MG}{\textsc{MG5}}
\newcommand*{\PY}{\textsc{PY}}

\begin{document}

\begin{titlepage}
  \pubblock

  \vfill
  \Title{Monte Carlo modeling in single top quark and in other rare top-related processes at ATLAS and CMS}
  \vfill
  \Author{Simon Berlendis,\\
    on behalf of the ATLAS and CMS Collaborations\support}
  \Address{\institute}
  \vfill

  \begin{Abstract}

    The production of top quarks through single or rare production modes has become important due to the large amount of data collected by both ATLAS and CMS at the LHC. Many searches are now studying these processes either as a targeted signal or as an important source of background. This document reviews the Monte Carlo simulations used to model these processes in ATLAS and CMS, and how the modeling systematic uncertainties are estimated. Many analyses have also recently released a large variety of unfolded distributions. These distributions are shown and the modeling from various Monte Carlo generators are compared to the data.

  \end{Abstract}

  \vfill
  \begin{Presented}
    $12^\mathrm{th}$ International Workshop on Top Quark Physics\\
    Beijing, China, September 22--27, 2019
  \end{Presented}
  \vfill
\end{titlepage}
\def\thefootnote{\fnsymbol{footnote}}
\setcounter{footnote}{0}

\section{Introduction}

With the nearly 150~\mbox{fb\(^{-1}\)} of data collected by ATLAS~\cite{ATLAS} and CMS~\cite{CMS} during Run~2 of the LHC, the production of top quarks through single or rare production modes has become important. Many searches for new physics are studying these processes either as a targeted signal or as an important source of background. For most of these analyses, these processes are simulated using Monte Carlo (MC) generators. Hence it is important to understand the validity and the accuracy of the modeling of these simulations.

This document reviews how well single top production and other rare top production modes are modeled. In this context, the detemination of the resulting systematic uncertainties of the corresponding measurements is discussed. Several unfolded distributions are shown and compared to the predictions from various MC generators. Section~\ref{sec:tchan} and~\ref{sec:tw} focus on single top-quark modeling through the $t$-channel and $tW$-channel production mode respectively, while Section~\ref{sec:ttV} reviews the modeling of the top-quark pair production associated with an additional vector boson (a $W/Z$-boson or a photon $\gamma$). 

\section{$t$-channel single top quark}
\label{sec:tchan}

\begin{figure}[p]
\centering
\includegraphics[height=1in]{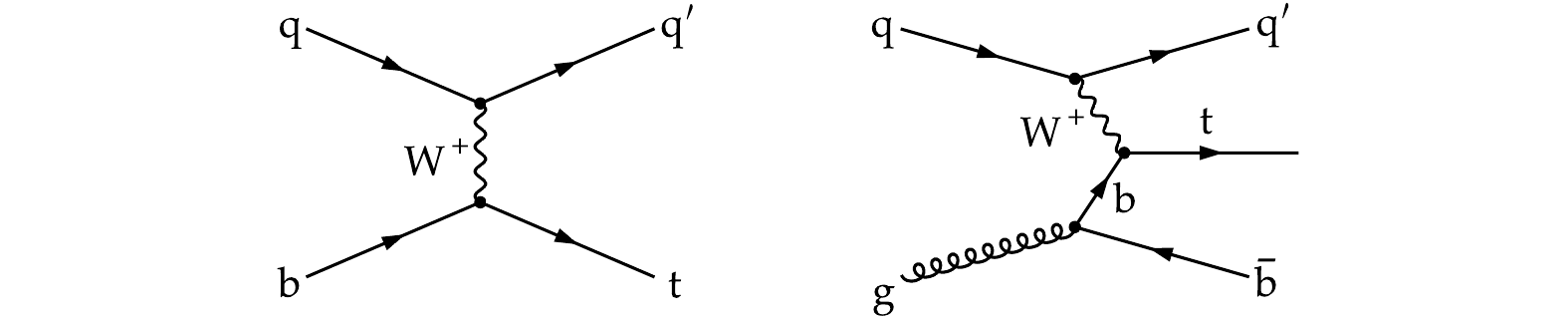}
\caption{Born-level Feynman diagrams for single top quark production in the $t$-channel (left) for the 5-flavor initial state and (right) for the 4-flavor initial state (figures taken from Ref~\cite{Sirunyan:2019hqb}).}
\label{fig:CMS-TOP-17-001}
\end{figure}

The $t$-channel, for which the Born-level Feynman diagrams are shown in Figure~\ref{fig:CMS-TOP-17-001}, is the single top-quark production mode that has the largest cross section. Its value has already been measured at 13~\TeV\ by both ATLAS and CMS~\cite{Sirunyan:2018rlu,Sirunyan:2019hqb,Aaboud:2016ymp}.  In CMS, this process is simulated at Next-Leading Order (NLO) using the matrix element (ME) generator \POWHEG~v2~\cite{Alioli:2010xd} matched with the parton shower generator \PYTHIA~8~\cite{Sjostrand:2014zea} (\PW2+\PY8), with the top-quark mass as the matching scale choice. NNPDF3.0~\cite{Ball:2014uwa} and NNPDF2.3~\cite{Watt:2012tq} are used as parton distribution functions (PDF) for the ME and PS, respectively. The 4-flavor scheme (4FS) shown in the right diagram in Figure~\ref{fig:CMS-TOP-17-001} is used as it has a more accurate modeling of the spectator $b$-quark than the 5-flavor scheme~\cite{Frederix:2012dh} (left diagram in Figure~\ref{fig:CMS-TOP-17-001}). The set of parameters in \PYTHIA\ are defined by the CUETP8M1 tune~\cite{Khachatryan:2015pea}. In ATLAS, this process is simulated at NLO using \POWHEG~v1+\PYTHIA~6 (\PW1+\PY6) using a 4FS and a scale set by $4\times\sqrt{m_b^2+p_{T,b}^2}$, where $m_b$ and $p_{T,b}$ are the mass and transverse momentum of the spectator $b$-quark. The PDF and the \PYTHIA\ parameters are defined using CT10f4~\cite{Dulat:2015mca} and Perugia2012 (P2012) tune parameter set~\cite{Skands:2010ak}.

\begin{figure}[p]
\centering
\includegraphics[height=2in]{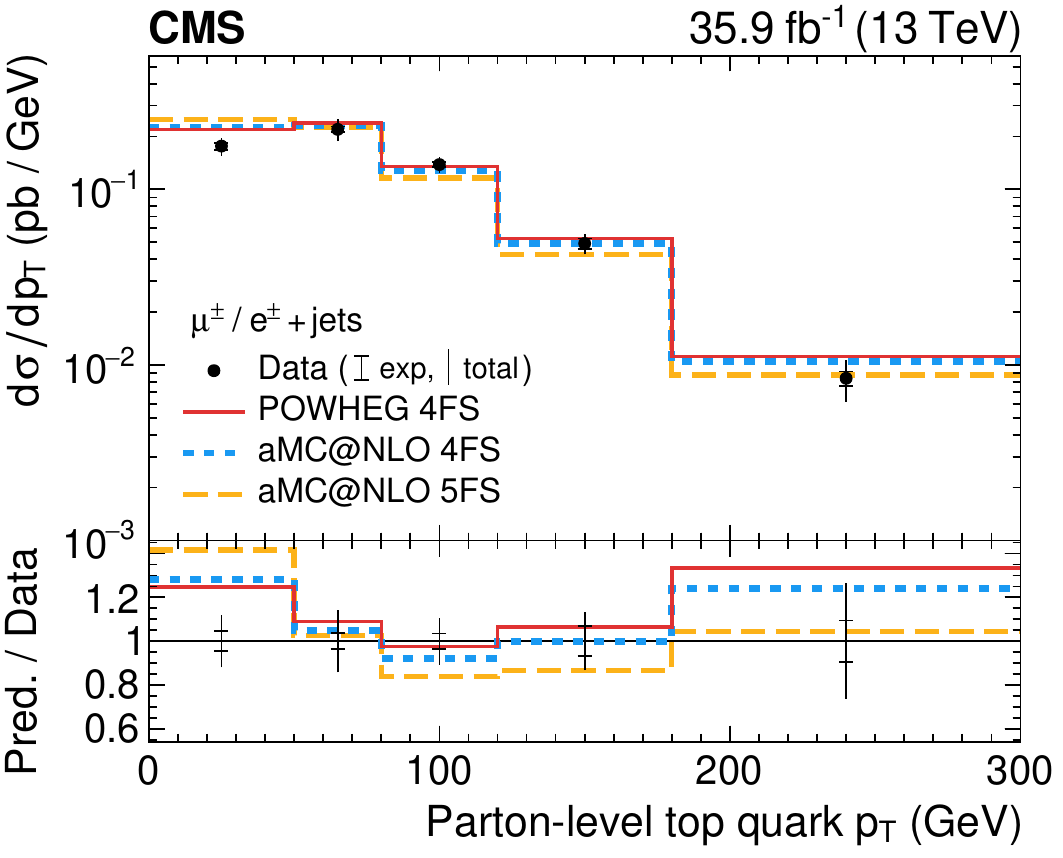}
\includegraphics[height=2in]{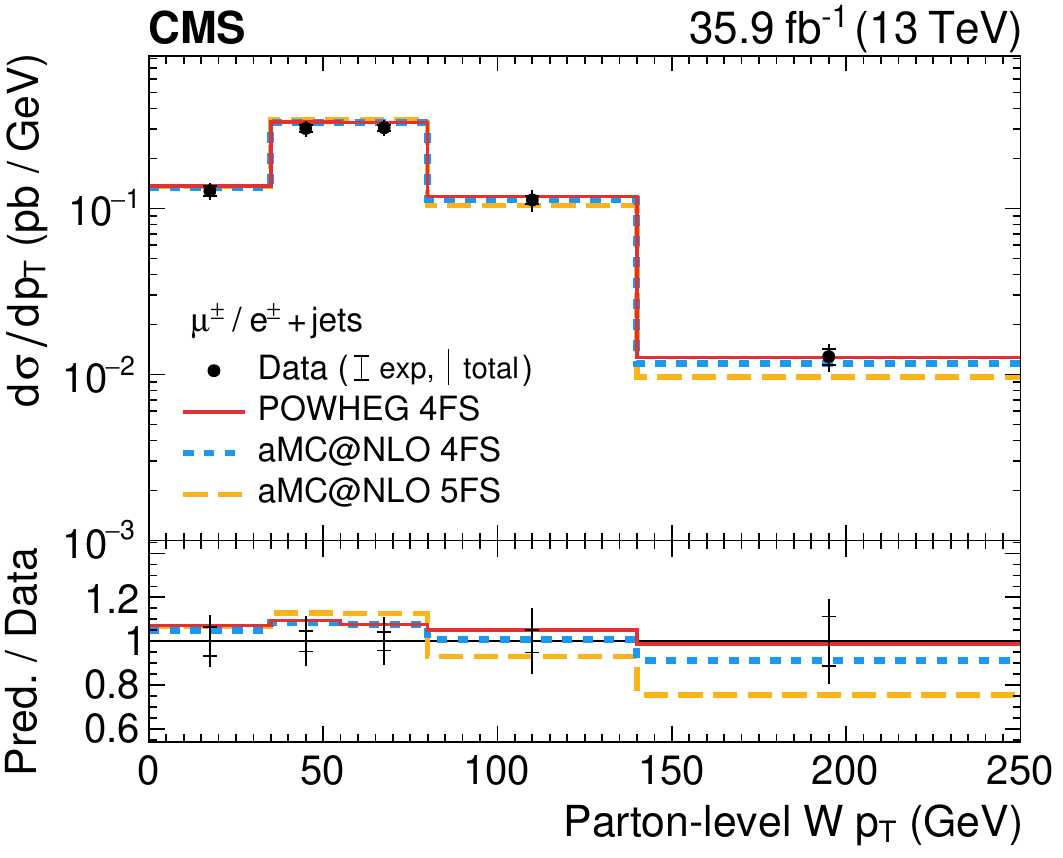}
\includegraphics[height=2in]{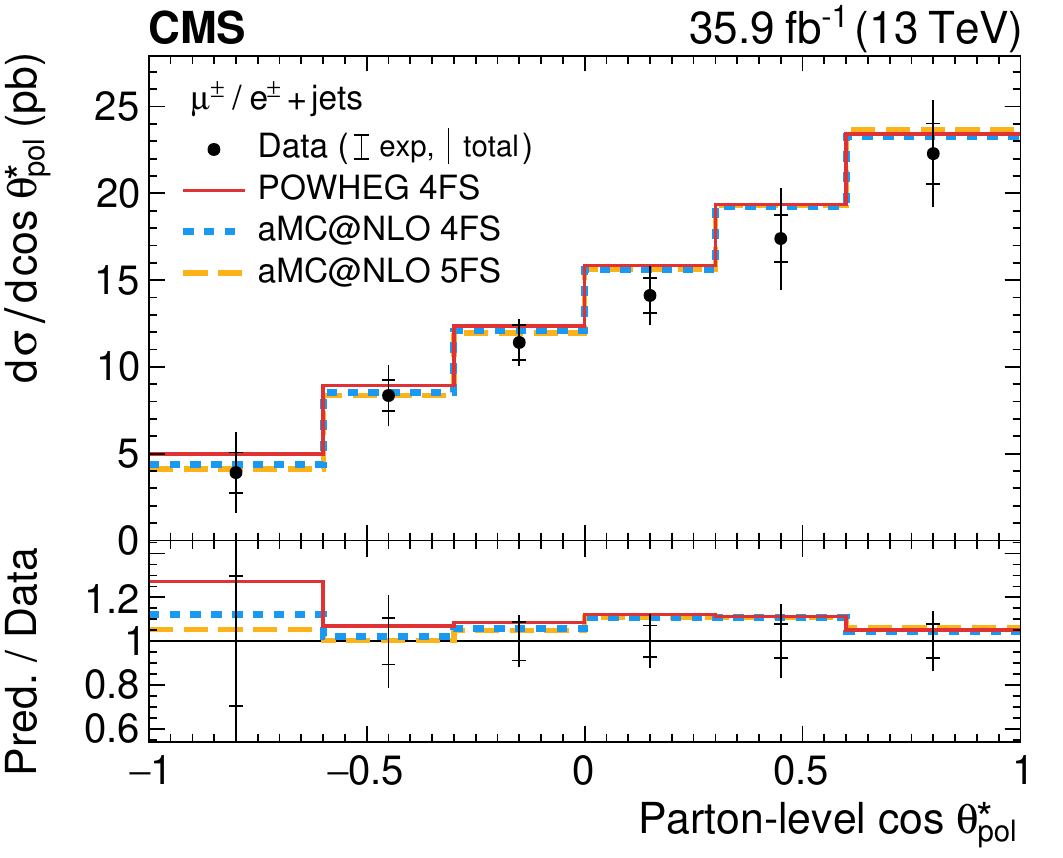}
\includegraphics[height=2in]{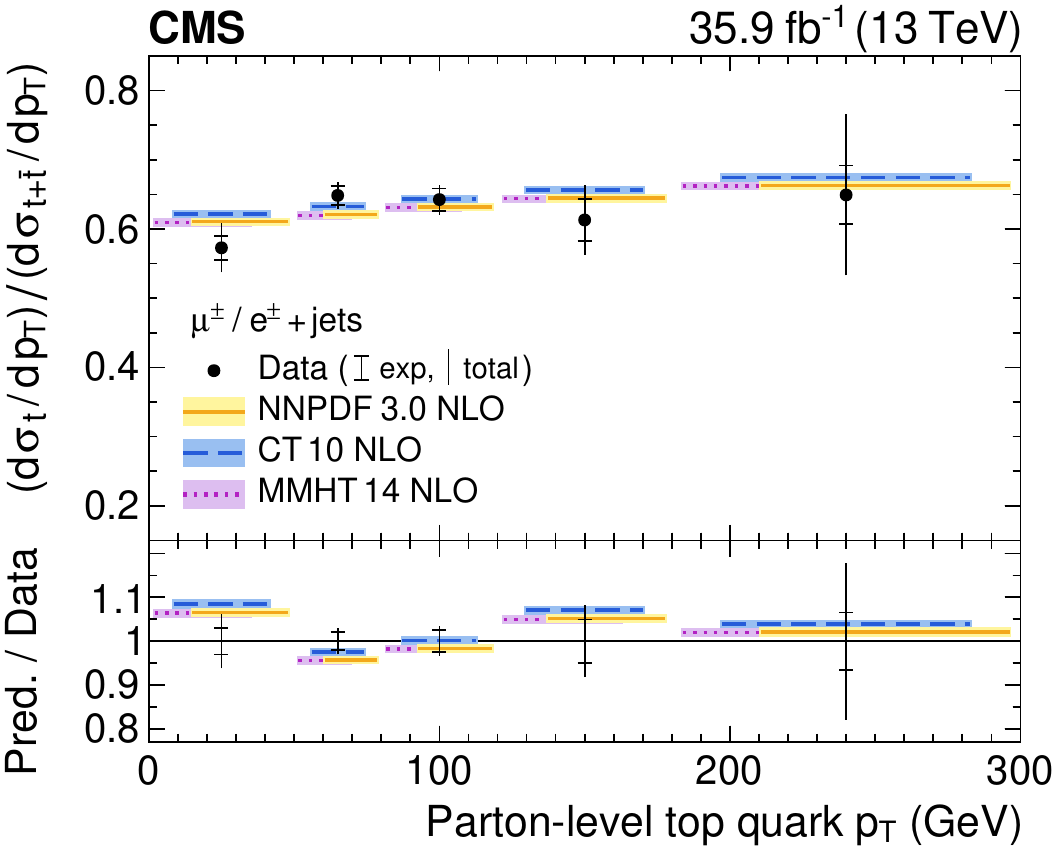}
\caption{Normalised $t$-channel differential cross sections unfolded from data with respect to (top left) the top quark and (top right) $W$-boson transverse momentum and (bottom left) the cosine of the top quark polarization angle. The ratio of the top quark to the sum of top quark and antiquark $t$-channel cross section as function of the top quark transverse momentum is shown on the bottom right Figure. All distributions are unfolded to the parton level (figures taken from Ref~\cite{Sirunyan:2019hqb}).}
\label{fig:CMS-TOP-17-023}
\end{figure}

In CMS, the modeling uncertainties coming from the ME and the PS are respectively estimated by varying the factorization and renormalization scale in the ME, and by varying the factorization scale in the PS. In ATLAS, these ME uncertainties are estimated by comparing predictions from \PW1+\PY6 with the \MadGraph~\cite{Alwall:2014hca} ME interfaced with the \PYTHIA~6\ PS (\MG+\PY6). Uncertainties introduced by PS modeling are estimated by comparing \PW1+\PY6 with \POWHEG~v1 interfaced with \HERWIG++~\cite{Bahr:2008pv} PS (\PW1+\Hpp). The uncertainty from the QCD radiation is estimated in ATLAS by varying the factorization and renormalization scale together with the tuned values from P2012. In CMS, it is estimated by varying the damping factor $h_{damp}$ in \POWHEG\ around its tuned value $h_{damp} = 1.581^{+0.658}_{-0.585}\times m_{t}$. Both experiments take into account the uncertainty from the PDF set. The PS uncertainty is the dominant modeling uncertainty for this process, with a relative uncertainty of 13\% for the measured $t$-channel cross section for both ATLAS and CMS. However, this uncertainty cancels out when considering the ratio of the top-quark production over the anti-top quark production.

Figure~\ref{fig:CMS-TOP-17-023} shows several $t$-channel unfolded distributions at parton level published by CMS~\cite{Sirunyan:2019hqb}. A reasonable agreement with data is observed for \POWHEG\ and \MadGraph\ predictions, except at low top-quark transverse momentum. The cosine of the top quark polarization angle, defined as the angle between the spectator quark and the lepton from top-quark decay in the top-quark rest frame, is very well modeled by both generators. The ratio of top quark to the sum of the top quark and antiquark cross section is also well predicted by the different NLO PDF sets.

\section{$tW$-channel single top quark}
\label{sec:tw}

\begin{figure}[p]
\centering
\includegraphics[height=1in]{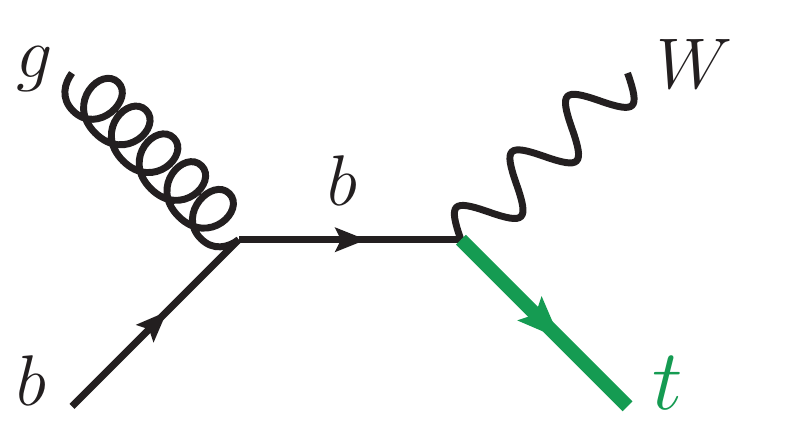}
\caption{Born-level Feynman diagrams for single top quark production in the $tW$-channel.}
\label{fig:tw}
\end{figure}

The cross section for the single top production in the $tW$-channel with the Born-level diagram shown in Figure~\ref{fig:tw}, was measured at at 13~\TeV\ by both ATLAS and CMS~\cite{Aaboud:2016lpj,Aaboud:2017qyi,Sirunyan:2018lcp}. In CMS, this process is simulated at NLO using \PW1+\PY8 using NNPDF3.0/NNPDF2.3 and the CUETP8M1 tune. In ATLAS, it is simulated at NLO using \PW1+\PY6 using CT10 and the P2012 tune. At NLO, a special treatment needs to be applied during the MC generation of the $tW$-channel in order to remove any overlap and interference with the top-quark pair ($t\bar{t}$) MC generation. Both experiments use the diagram removal (DR) scheme~\cite{Demartin:2016axk} for their nominal $tW$-channel MC generations.

\begin{figure}[p]
\centering
\includegraphics[height=2.5in]{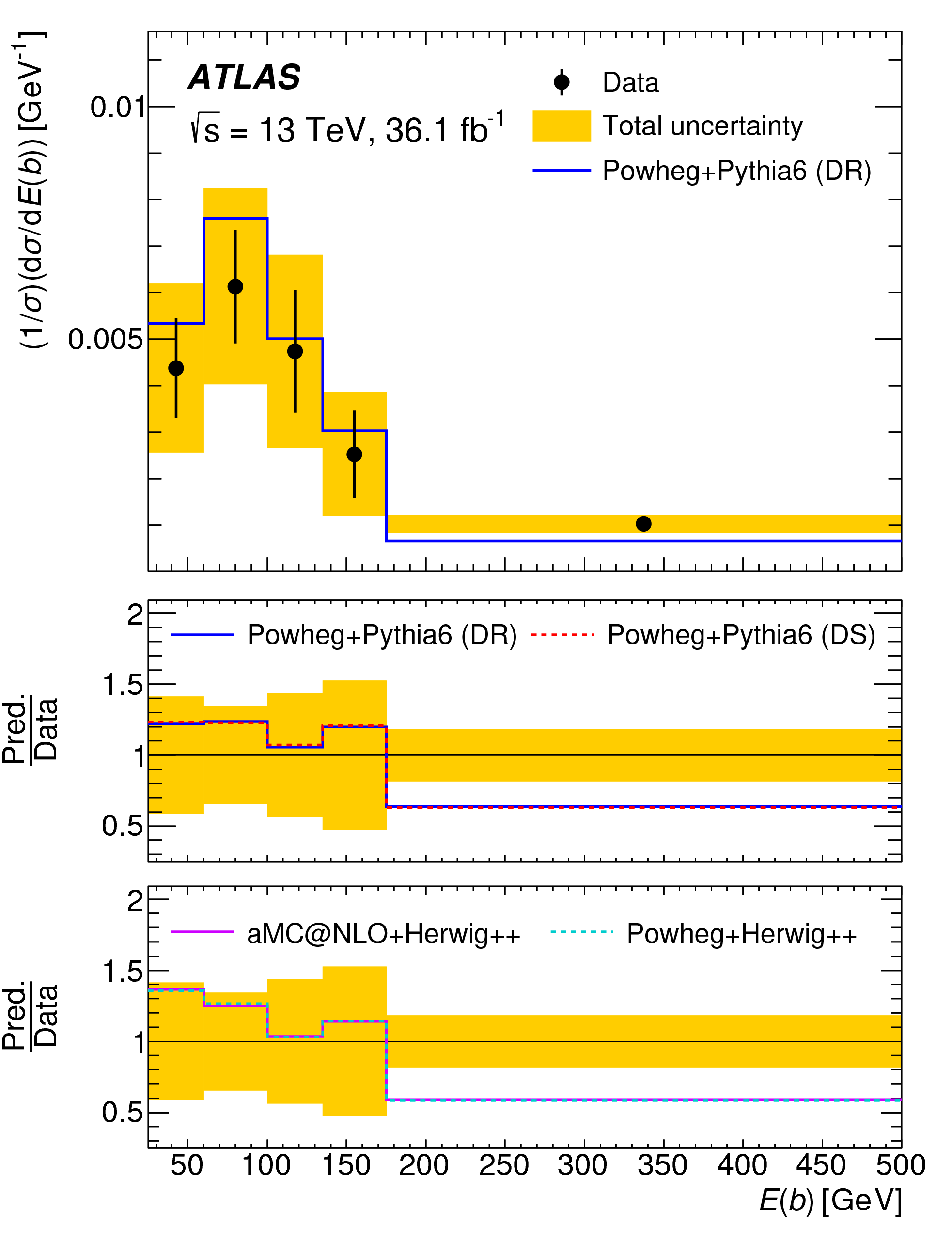}
\includegraphics[height=2.5in]{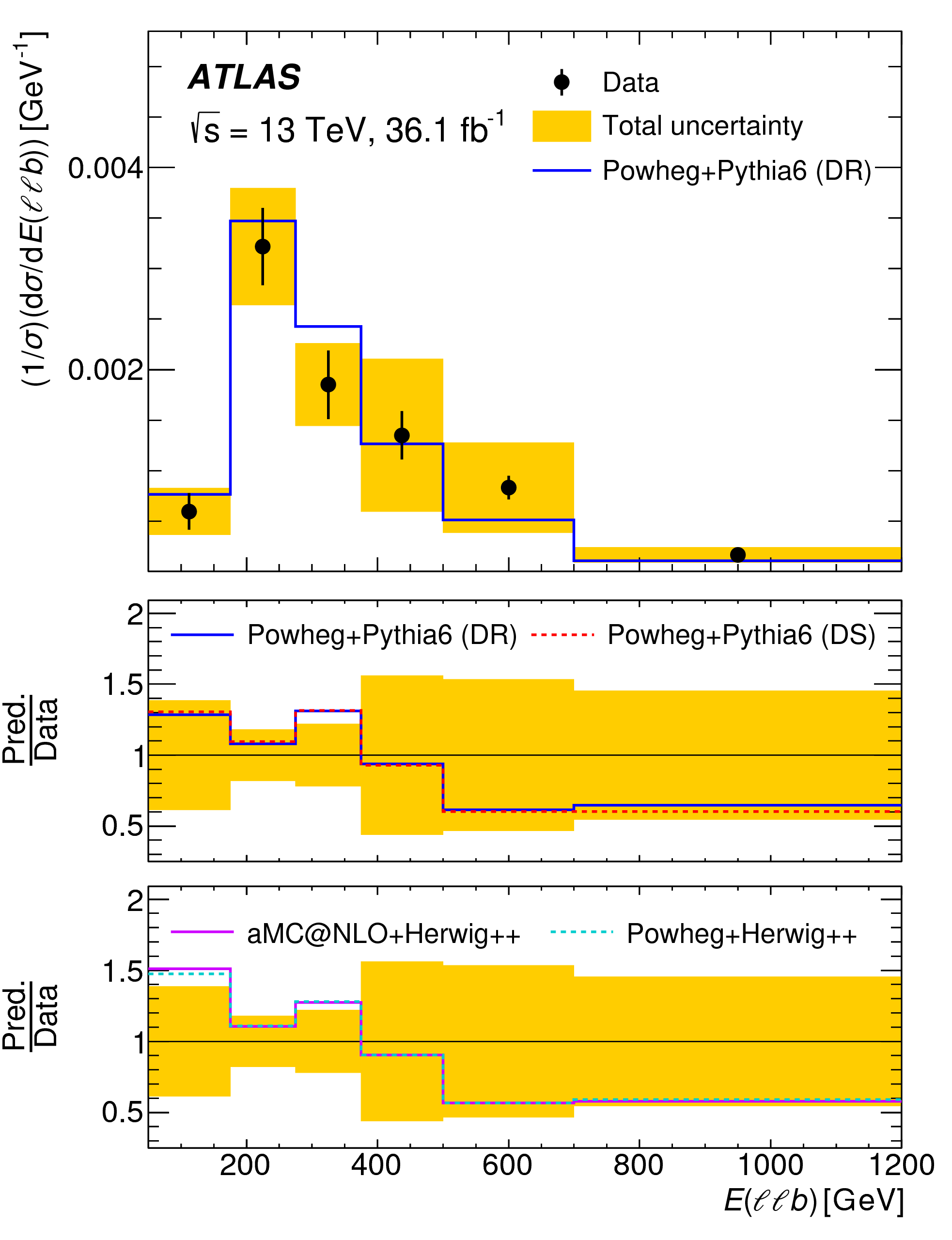}
\includegraphics[height=2.5in]{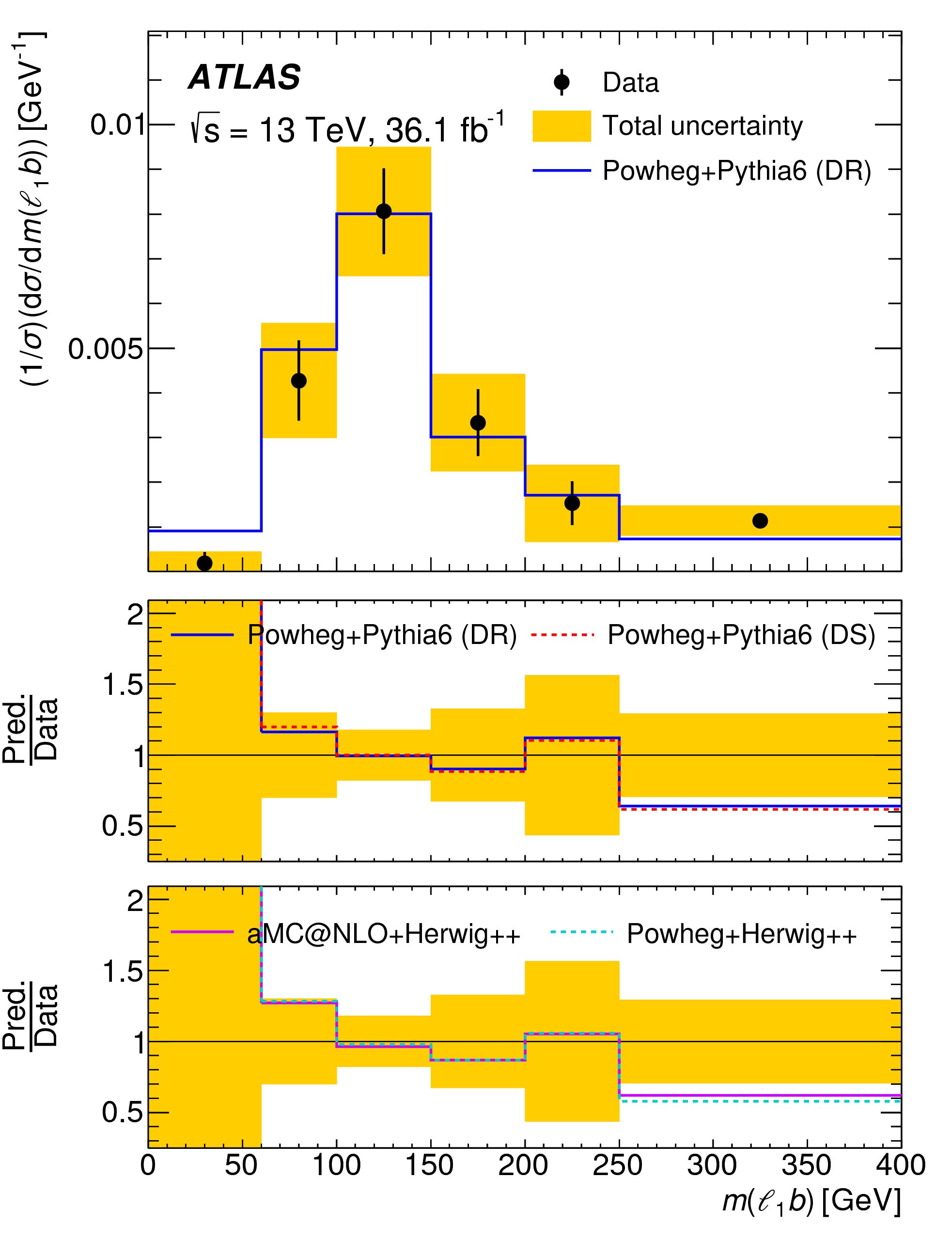}\\
\includegraphics[height=2.5in]{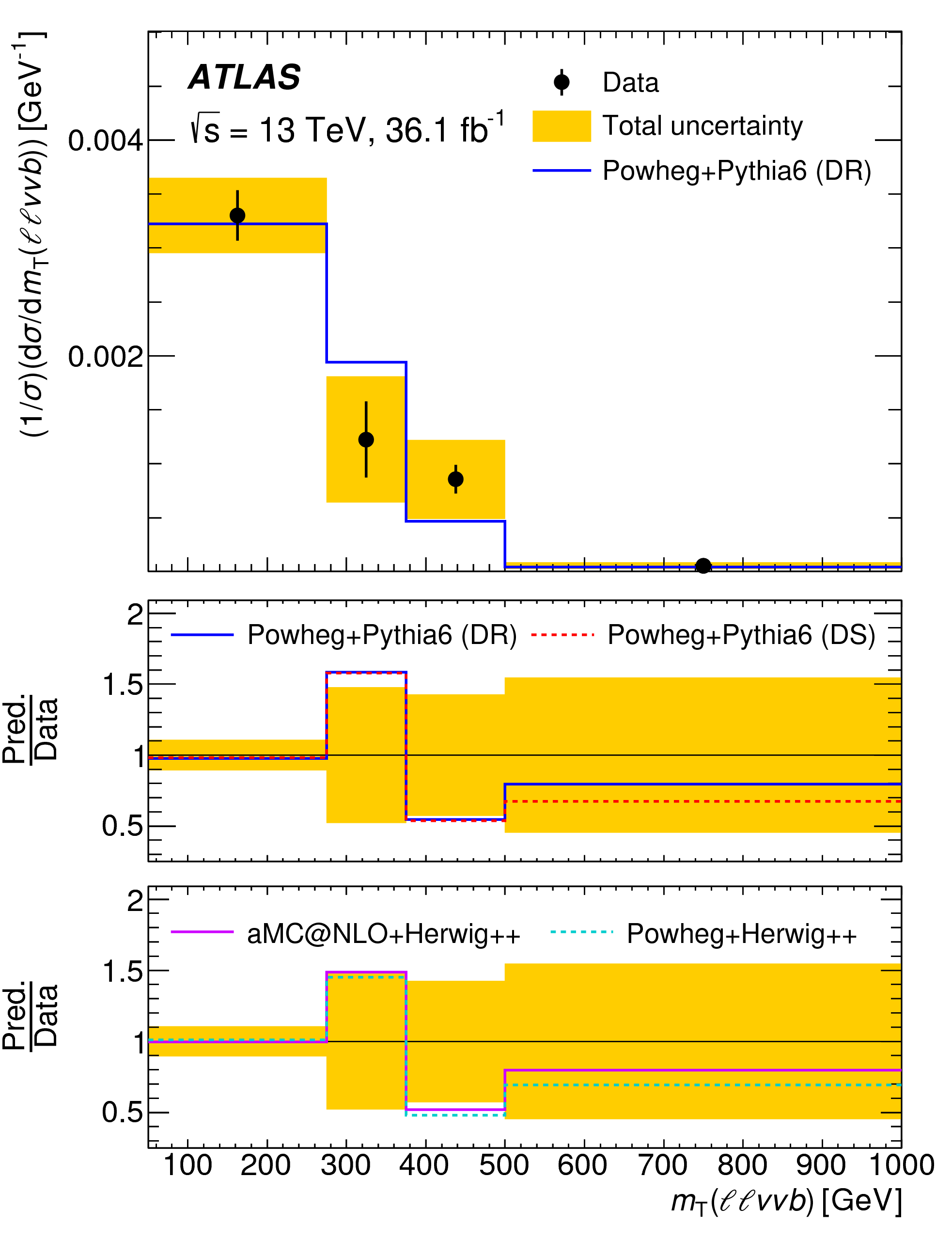}
\includegraphics[height=2.5in]{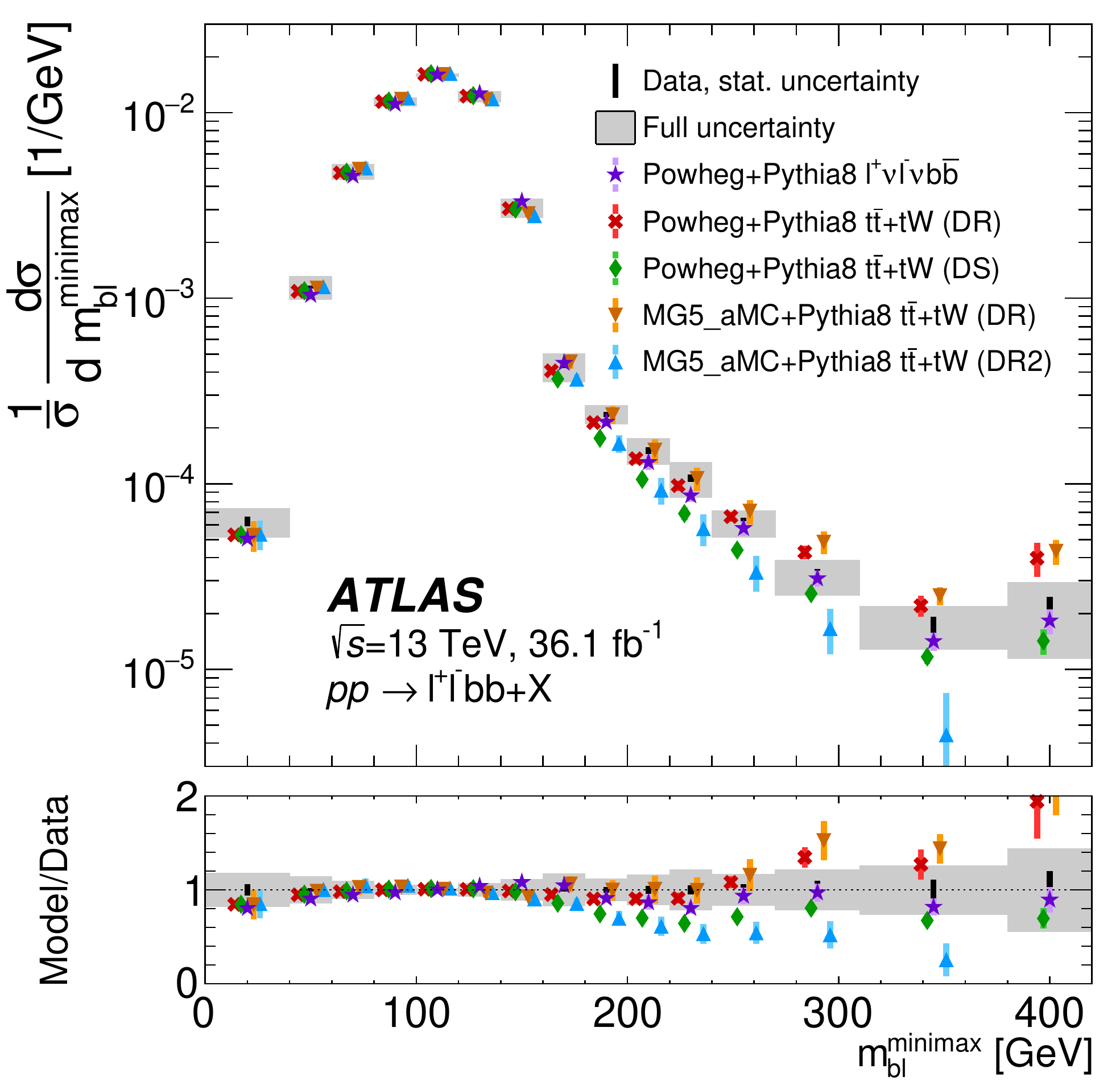}
\caption{Normalised $tW$-channel differential cross sections unfolded from data with respect to the energy of (top left) the $b$-quark and (top middle) the lepton+$b$-quark system, to (top right) the invariant mass of two leptons + $b$-quark system, to (bottom left) the transverse mass of all the final particles system and to (bottom right) the $m_{bl}^{\mathrm{minimax}}$ observable (figures taken from Ref~\cite{Aaboud:2017qyi} and \cite{Aaboud:2018bir}).}
\label{fig:TOPQ-2016-12}
\end{figure}

Both experiments use the strategy described in Section~\ref{sec:tchan} to estimate the systematic uncertainties from the modeling of the $tW$-channel. Additional uncertainties on the parton shower related to the color reconnection, the $b$-quark fragmentation and the $B$-hadron branching ratio are considered by CMS. Both experiments consider also an additional uncertainty on the choice of $tW$-$t\bar{t}$ overlap removal scheme by comparing the predictions between the DR scheme and the diagram subtraction (DS) scheme~\cite{Demartin:2016axk}. The impact of the modeling uncertainty on the measured $tW$-channel total cross section is significantly different between ATLAS and CMS. While the uncertainty from modeling is below 3\% in CMS~\cite{Sirunyan:2018lcp}, ATLAS quotes a contribution of about 18\% (7\%) introduced by the choice of ME (PS) models~\cite{Aaboud:2016lpj}.

Figure~\ref{fig:TOPQ-2016-12} shows several $tW$-channel unfolded distributions at particle level in the 2$\ell$+1$b$ event region published by ATLAS~\cite{Aaboud:2017qyi}, compared with several MC predictions from \PW1+\PY6 using the DR or the DS scheme and from \PW1+\Hpp\ and \MG+\Hpp. In general, a fair agreement with data is seen with all MC predictions. Figure~\ref{fig:TOPQ-2016-12} shows the $tW$+$t\bar{t}$ unfolded distributions in the 2$\ell$+2$b$ event region compared to several MC predictions that use different $tW$-$t\bar{t}$ overlap removal schemes~\cite{Aaboud:2018bir}. The variable used for this distribution, referred as $m_{bl}^{\mathrm{minimax}}$\footnote{$m_{bl}^{\mathrm{minimax}}=\min\{\max\{m_{b_1,\ell_1},m_{b_2,\ell_2}\},\max\{m_{b_1,\ell_2},m_{b_2,\ell_1}\}\}$}, is designed to be sensitive to the interference between $tW$ and $t\bar{t}$ at high values. The \PW2+\PY8 $pp\rightarrow \ell^+\nu\ell^-\nu b\bar{b}$ predictions where both $tW$ and $t\bar{t}$ are generated give the best agreement with data. The \PW2+\PY8 $tW$ samples combined with the $t\bar{t}$ sample with the same generator, using the DR or the DS scheme give a relatively fair agreement with data, and the difference between both predictions gives an appropriate systematic uncertainty on the choice of the scheme. The \MG+\PY8 prediction using the alternative scheme called DR2~\cite{Demartin:2016axk} shows however some tension with the data on the tail of the distribution. 

\section{Top quark pairs associated with a vector boson}
\label{sec:ttV}

\begin{figure}[p]
\centering
\includegraphics[height=1.5in]{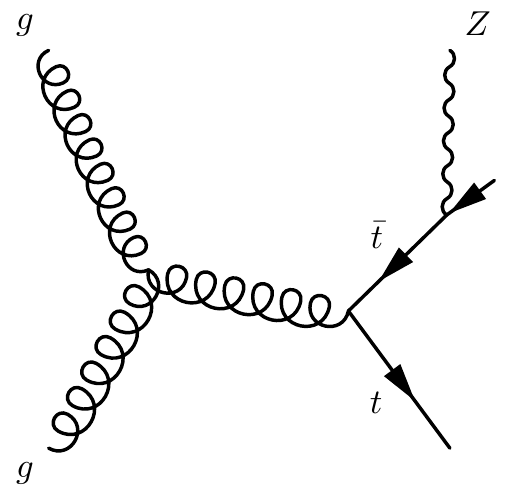}
\includegraphics[height=1.5in]{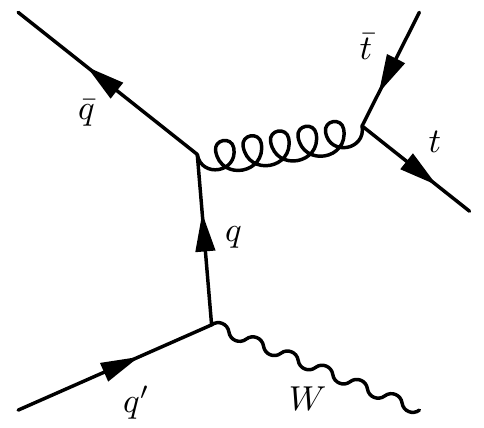}
\includegraphics[height=1.5in]{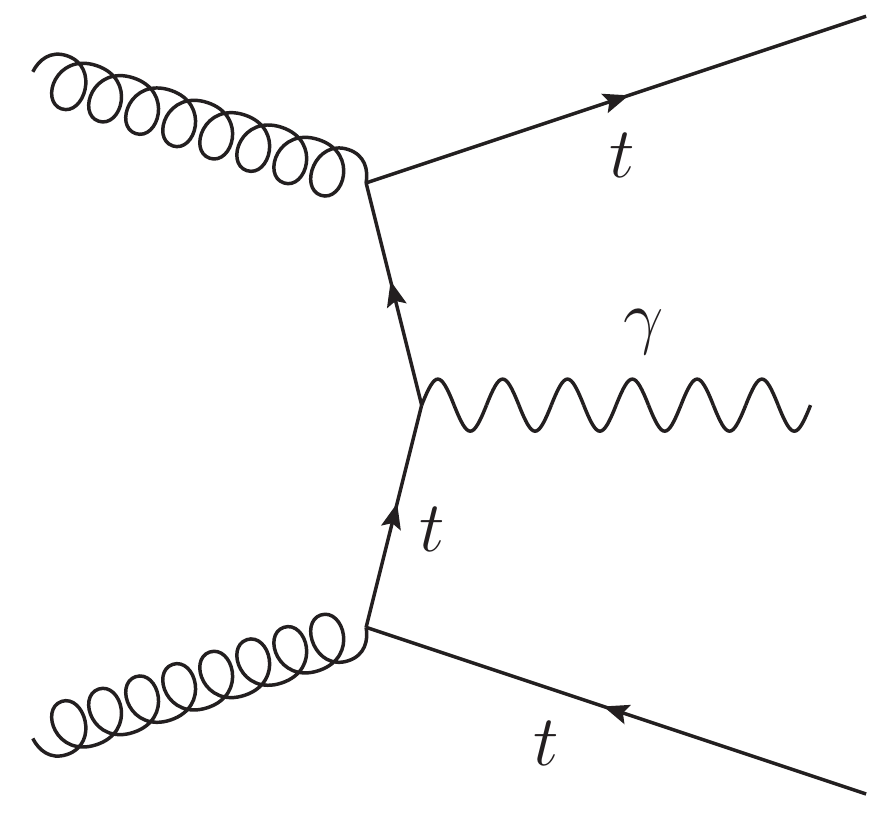}
\caption{Born-level Feynman diagrams for the production of top quark pairs associated with an additional (left) $Z$-boson, (middle) $W$-boson or (right) photon $\gamma$.}
\label{fig:ttv}
\end{figure}

\begin{figure}[p]
\centering
\includegraphics[height=4.5in]{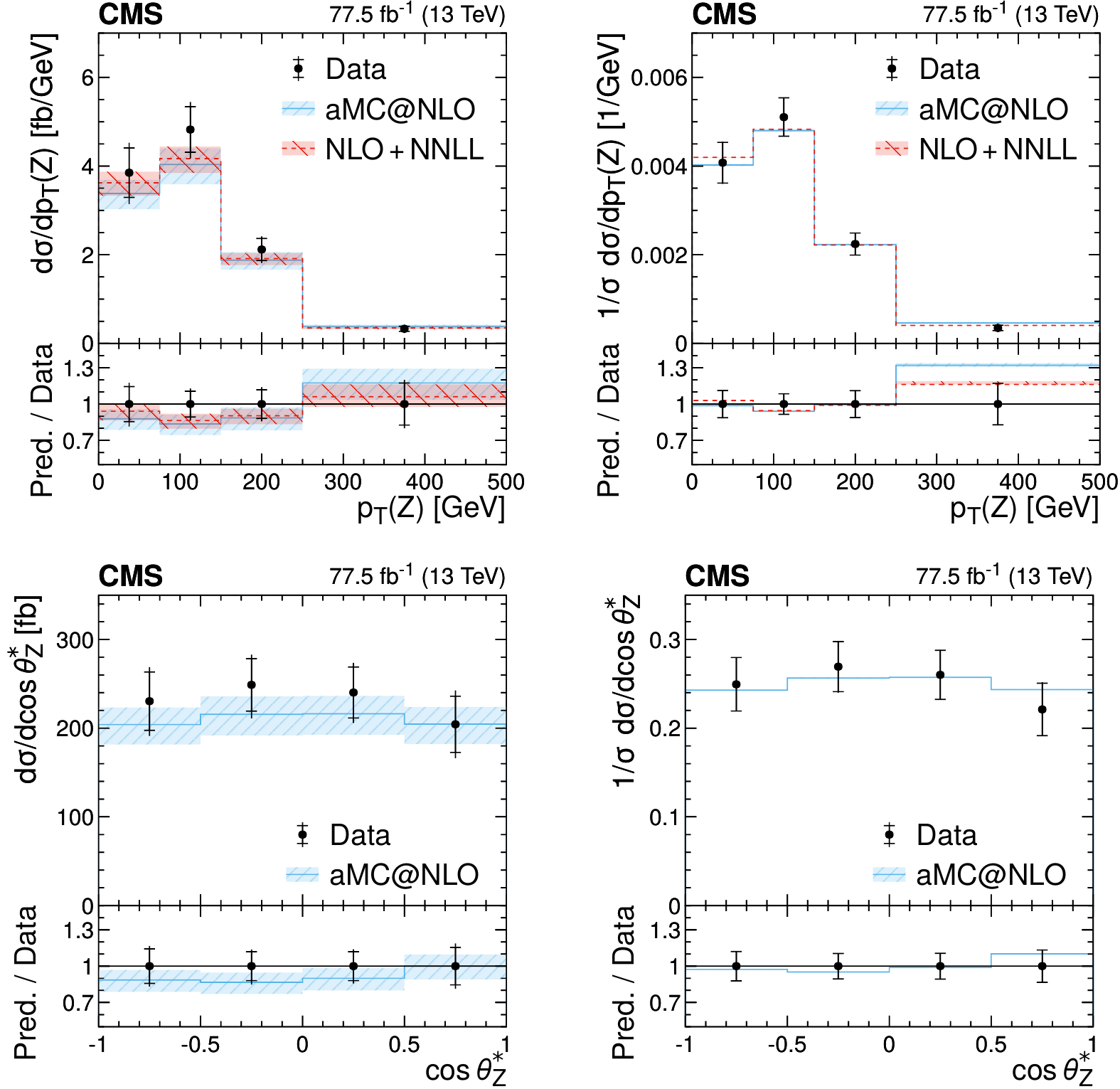}
\caption{(left) Unnormalised and (right) normalised $t\bar{t}Z$ differential cross sections unfolded from data at particle level with respect to (top) the $Z$-boson transverse momentum and (bottom) the cosine between the $Z$-boson in the detector frame and the negative lepton in the $Z$-boson rest frame (figures taken from Ref~\cite{CMS:2019too}).}
\label{fig:CMS-TOP-18-009}
\end{figure}

Figure~\ref{fig:ttv} shows the Feynman diagrams for the production of top quark pairs associated with an additional $Z/W$ boson (referred as $t\bar{t}Z$ and $t\bar{t}W$). The $t\bar{t}Z$ cross section was measured by both ATLAS and CMS~\cite{Aaboud:2019njj,CMS:2019too}, while the  $t\bar{t}W$ cross section was only measured by ATLAS at 13~\TeV~\cite{Aaboud:2019njj}. For both experiments, these processes are simulated at NLO using \MG+\PY8. The NNPDF2.3 PDF set and A14~\cite{ATL-PHYS-PUB-2014-021} \PYTHIA\ tune is used by ATLAS, while CMS uses NNPDF3.0/NNPDF2.3 (NNPDF3.1~\cite{Ball:2017nwa}) for the ME/PS PDF set and the CUETP8M1 (CP5) \PYTHIA\ tune for the 2015--2016 (2017--2018) datasets.
Both experiments use the factorization and renormalization scale variations as a theoretical modeling uncertainty. The uncertainty from the PS is estimated in CMS by varying the factorization scale in the initial and final state radiation, as well as by comparing different color reconnection models. In ATLAS, a comparison between \MG+\PY8 predictions with the predictions from a \SHERPA\ sample is used as an uncertainty on the choice of the MC generator. The variations of the A14 PS tune is also used as an uncertainty on the PS modeling. The uncertainty from the PDF set is taken into account by both experiments. The total modeling uncertainty on the $t\bar{t}Z$ and $t\bar{t}W$ total cross section is estimated by ATLAS to be around 4.9\% and 8.5\%, respectively. CMS estimates each modeling uncertainty of up to 1\% of the measured cross section.
Figure~\ref{fig:CMS-TOP-18-009} shows the $t\bar{t}Z$ unfolded distributions at the particle level in the 3$\ell$ event region published by CMS~\cite{Sirunyan:2019hqb}. Both the \MG+\PY8 prediction and the NLO theoretical prediction corrected at the Next-to-Next Leading Log (NNLL)~\cite{Kulesza:2019adl} are in good agreement with data. 

\begin{figure}[p]
\centering
\includegraphics[height=2.5in]{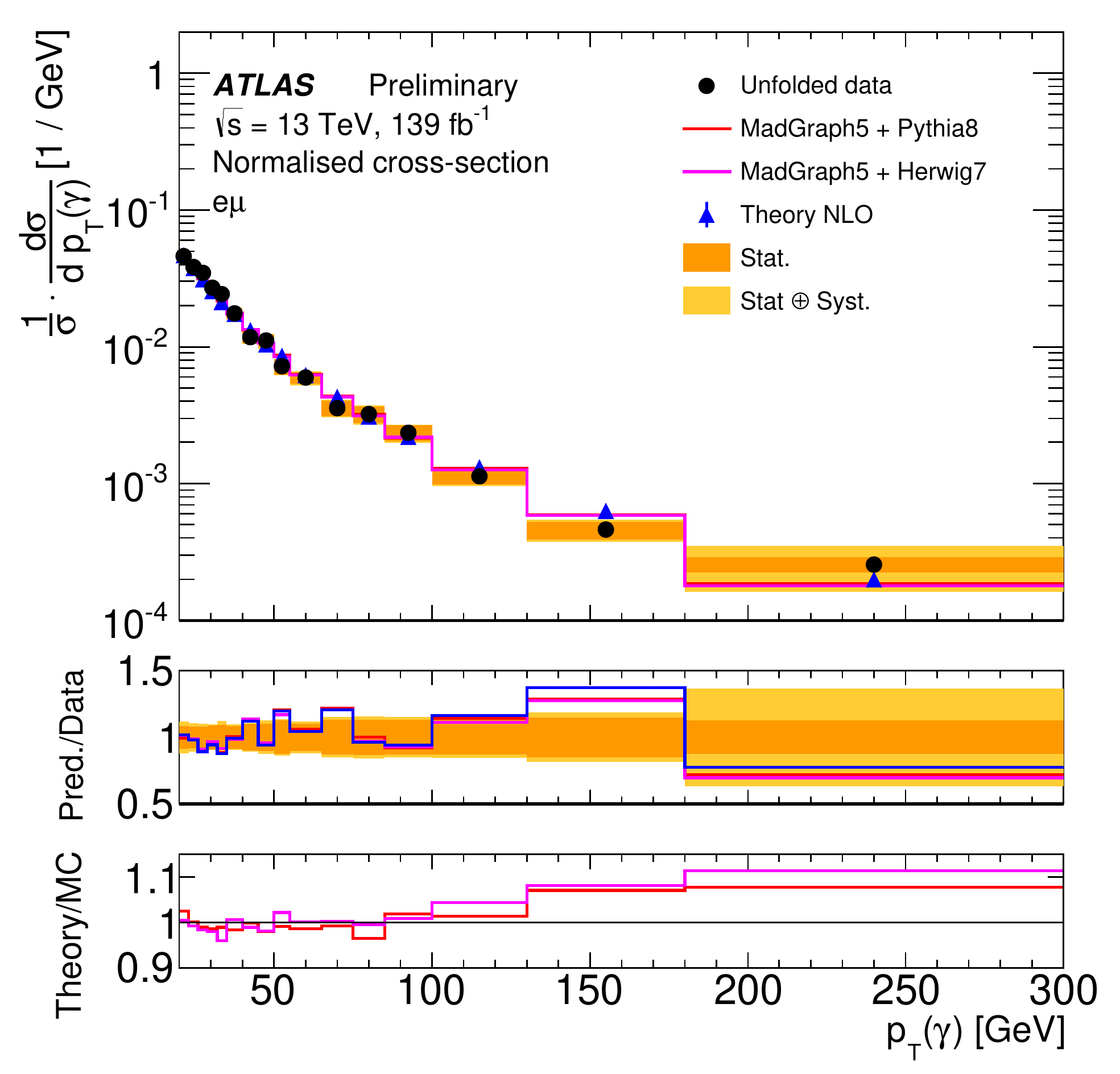}
\includegraphics[height=2.5in]{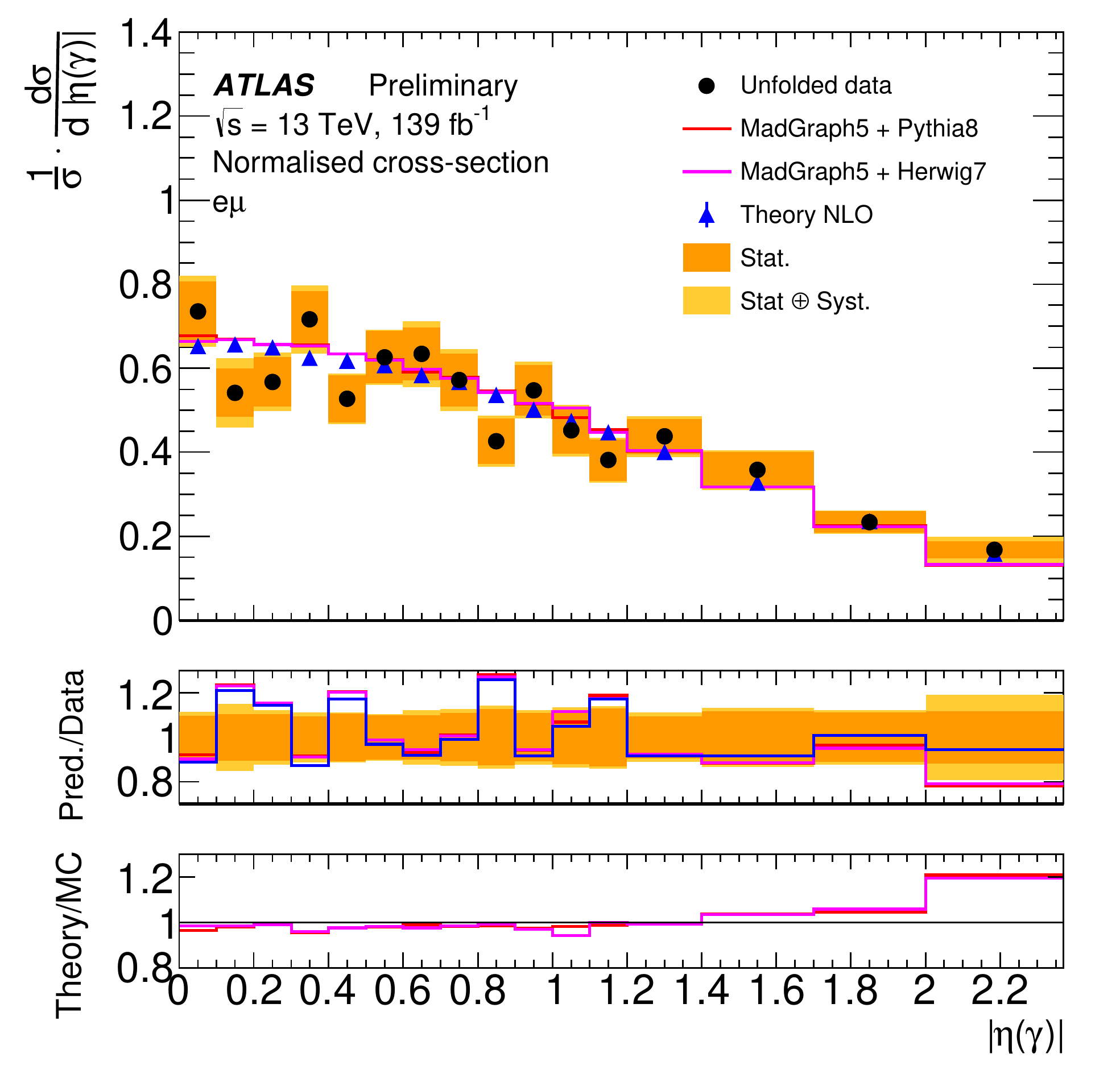}
\includegraphics[height=2.5in]{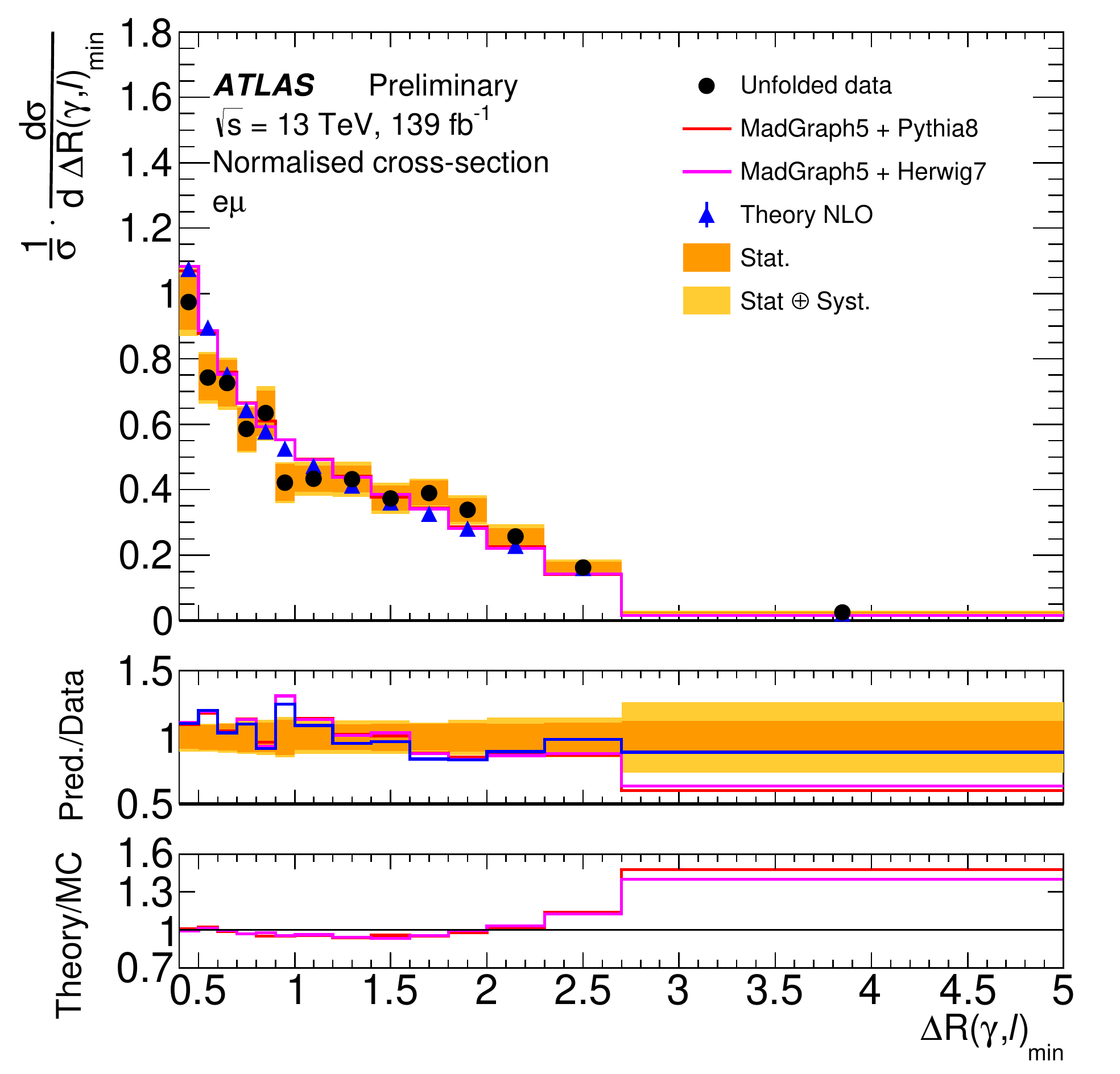}
\includegraphics[height=2.5in]{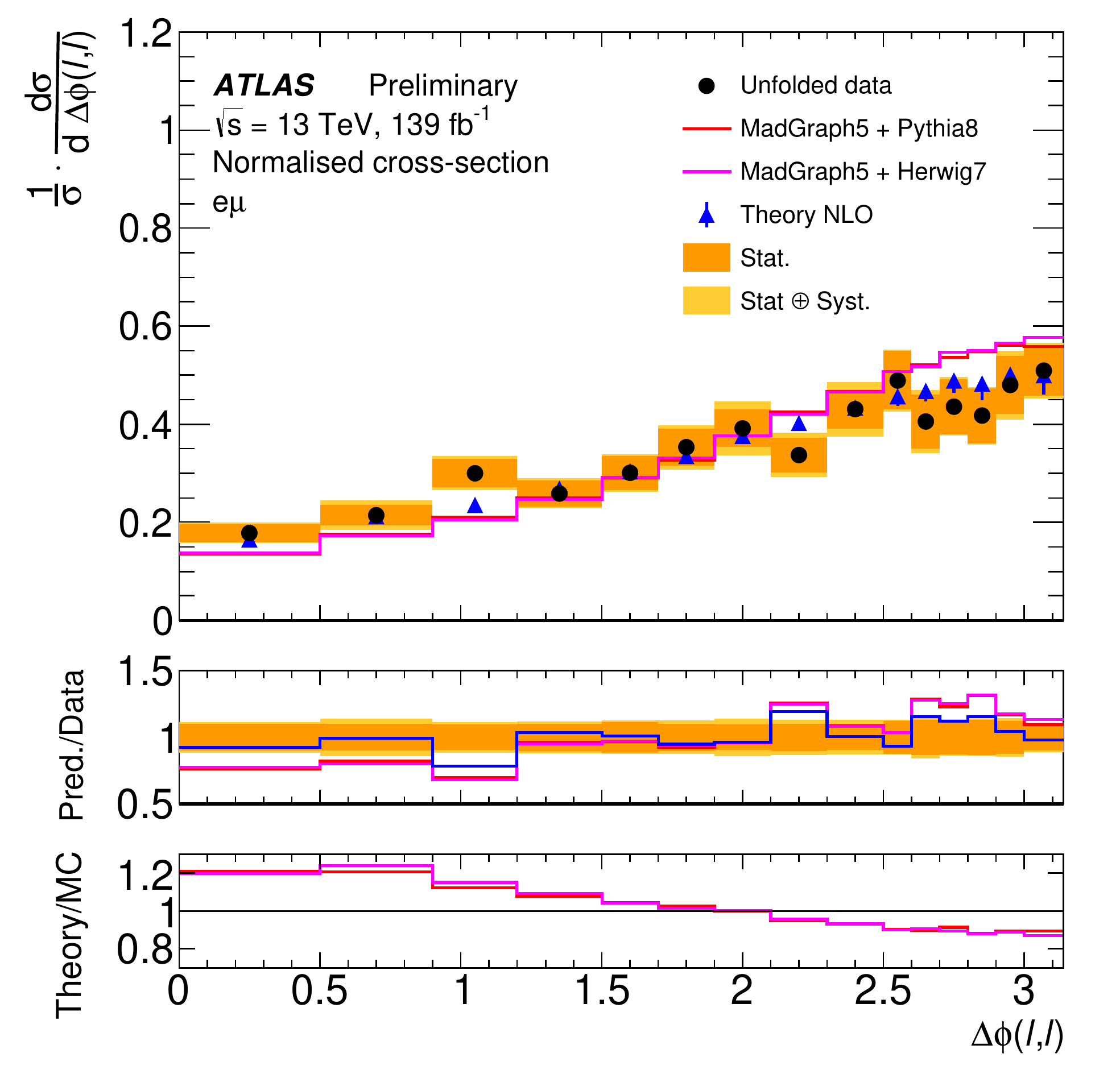}
\caption{Normalised $t\bar{t}\gamma$ differential cross sections unfolded from data at parton level with respect to (top left) the transverse momentum and (top right) the rapidity of the photon, to (bottom left) the minimum distance between the photon and the lepton and to (bottom right) the azimuthal angle between both leptons (figures taken from Ref~\cite{ATLAS-CONF-2019-042}).}
\label{fig:ATLAS-CONF-2019-042}
\end{figure}

The production of top-quark pairs in association with a photon (referred as $t\bar{t}\gamma$) has multiple production modes. The photon can arise from top-quark radiation, as shown in the left diagram of Figure~\ref{fig:ttv}, from the initial state radiation or as the radiation from charged top-quark decay products. This process has been measured by ATLAS at 13~\TeV~\cite{ATLAS-CONF-2019-042}. The associated production is simulated at Leading Order (LO) using \MG+\PY8 as a $pp\rightarrow(b\ell\nu)(b\ell\nu)\gamma$ process using the NNPDF2.3 PDF set and the A14 tune. The contribution from charged top-quark decay products is reduced by requiring the angular distance $\Delta R$ between the photon and any stable final particle to be greater than 0.2, and by requiring the transverse momentum of the photon to be greater than 15~\GeV. The rest of the phase-space is covered by a $t\bar{t}$ MC sample where the photon is generated by the PS. The production of $tW$-channel single top quark associated with a photon is also generated using a similar setup with the DR scheme applied.
The modeling uncertainty is estimated by varying the factorization and renormalization scale, by comparing the predictions between \MG+\PY8 and \SHERPA\ and between \MG+\PY8 and \MadGraph\ interfaced with \HERWIG~7 (\MG+\HE7), by varying the A14 tuned parameters and by taking the different NNPDF variations. The resulting total modeling relative uncertainty on the cross section measurement is around 3.4\%.
Figure~\ref{fig:ATLAS-CONF-2019-042} shows $t\bar{t}\gamma$ unfolded distributions at parton level in the electron-muon channel. The \MG+\PY8 and \MG+\HE7 predications are in general in good agreement with data, except in the distribution of the azimuthal angle between the two leptons. On the latter, the NLO theoretical prediction taken from Ref~\cite{Bevilacqua:2018woc} shows a better agreement with the data.

\section{Conclusion}

ATLAS and CMS have different approaches for MC modeling of the production of single top quark and top quark pair associated with a vector boson. Although the nominal MC samples use a similar generator setup, the modeling systematic uncertainties are estimated differently. While ATLAS uses the comparison between different MC generators, CMS uses the variations of the nominal generator scales.

Both experiments have however published a large variety of unfolded distributions associated with these processes, which opens the possibility of improving the MC modeling. A more realistic estimation of the modeling uncertainty can be assessed, and better MC tunes can be derived by taking into account the effect of the different processes.

New ATLAS and CMS searches will be published using the full Run~2 data from the LHC, showing more unfolded distributions related to new rare top-related processes. This should provide more information on the overall MC modeling in top physics in ATLAS and CMS.

\end{document}